\documentclass[aps,twocolumn,showlabels,showrefs,amsmath,amssymb,prl,superscriptaddress,floatfix,colors, nofootinbib]{revtex4}

\usepackage{lineno}
\usepackage{graphicx}
\usepackage{dcolumn}
\usepackage{bm}

\usepackage{graphicx}
\usepackage{dcolumn}
\usepackage{bm}
\usepackage{amssymb}
\usepackage{hyperref}
\usepackage{multirow}
\usepackage{color}
\usepackage{cancel}
\usepackage[normalem]{ulem}

\newcommand{\ylann}[1]{{\color{black} #1}}

\usepackage[cp1251]{inputenc}

\makeatletter

\newcommand*{\sumcirclearrowleft}{%
 \DOTSB
 \mathop{
  \mathchoice
   {\rlap{\kern.25em\rotatebox[origin=c]{-90}{$\circlearrowleft$}}{\sum}}
   {\vcenter{\rlap{\kern.2em\rotatebox[origin=c]{-90}{$\scriptscriptstyle\circlearrowleft$}}}{\sum}}
   {\sum}{\sum}
 }\slimits@
}

\newcommand*{\sumcirclearrowright}{%
 \DOTSB
 \mathop{
  \mathchoice
   {\rlap{\kern.25em\rotatebox[origin=c]{90}{$\circlearrowright$}}{\sum}}
   {\vcenter{\rlap{\kern.2em\rotatebox[origin=c]{90}{$\scriptscriptstyle\circlearrowright$}}}{\sum}}
   {\sum}{\sum}
 }\slimits@
}

\makeatother

\begin{document}

\title{Activity leads to topological phase transition \\in  2D populations of heterogeneous oscillators}


\author{Ylann Rouzaire\textsuperscript{\textdagger}} 
\email{rouzaire.ylann@gmail.com}
 \affiliation{Departament de F\'isica de la Materia Condensada, Universitat de Barcelona, Mart\'i i Franqu\`es 1, E08028 Barcelona, Spain}
  \affiliation{UBICS University of Barcelona Institute of Complex Systems, Mart\'i i Franqu\`es 1, E08028 Barcelona, Spain}
 \affiliation{Computing and Understanding Collective Action (CUCA) Lab, Mart\'i i Franqu\`es 1, E08028 Barcelona, Spain}%

  \author{Parisa Rahmani\textsuperscript{\textdagger}}
\affiliation{Laboratoire de Physique Th{\'e}orique et Mod{\'e}lisation, UMR 8089, CY Cergy Paris Universit{\'e}, 95302 Cergy-Pontoise, France}
 
 \author{Ignacio Pagonabarraga} 
 \affiliation{Departament de F\'isica de la Materia Condensada, Universitat de Barcelona, Mart\'i i Franqu\`es 1, E08028 Barcelona, Spain}
  \affiliation{UBICS University of Barcelona Institute of Complex Systems, Mart\'i i Franqu\`es 1, E08028 Barcelona, Spain}

 \author{Fernando Peruani} 
\affiliation{Laboratoire de Physique Th{\'e}orique et Mod{\'e}lisation, UMR 8089, CY Cergy Paris Universit{\'e}, 95302 Cergy-Pontoise, France}

 \author{Demian Levis} 
 \affiliation{Departament de F\'isica de la Materia Condensada, Universitat de Barcelona, Mart\'i i Franqu\`es 1, E08028 Barcelona, Spain}
 \affiliation{UBICS University of Barcelona Institute of Complex Systems, Mart\'i i Franqu\`es 1, E08028 Barcelona, Spain}
 \affiliation{Computing and Understanding Collective Action (CUCA) Lab, Mart\'i i Franqu\`es 1, E08028 Barcelona, Spain}%

\date{\today}
\begin{abstract}
Populations of heterogeneous, noisy oscillators on a two-dimensional lattice display short-range order. 
Here, we show that if the oscillators are allowed to actively move in space, the system undergoes instead a Berezenskii-Kosterlitz-Thouless transition 
and exhibits quasi-long-range order. 
This fundamental result connects two paradigmatic models -- XY and Kuramoto model --  and provides insight on the emergence of order in active systems. 
\end{abstract}
\maketitle
\def\thefootnote{\textdagger}\footnotetext{These authors contributed equally to this work.}


At the interface between statistical mechanics and dynamical systems there are 
two fundamental, strongly related models: the XY 
 and Kuramoto model. 
The XY (planar) model, with short-range interactions, is one of the corner stones of statistical mechanics. 
It plays a major role to understand symmetry breaking 
in systems with continuous symmetry and finds applications in  
boson gases \cite{DalibardBKT}, superconductors \cite{Hebard1980}, physics of melting \cite{Halperin1978}, and liquid crystals \cite{SinghRev}, among many other examples. 
The classical XY model can be expressed in terms of Langevin equation, in the over-damped limit and in the canonical ensemble, as   
\begin{equation}
   \gamma \dot{\theta}_i =  \sigma\, \omega_i +   {J} \sum\limits_{j=1}^{N}{a_{ij}} \sin(\theta_j - \theta_i) + \sqrt{2\gamma\,T\,}\eta_i \, ,
    \label{eq:xymodel}
\end{equation}
where  ${\theta}_i$ is the 
 phase of the $i$th spin or rotor,  $a_{ij}=1$ if $j$ is a nearest neighbor of $i$ on a regular lattice of dimension $d$ ($a_{ij}=0$ otherwise), 
$\gamma$ is the damping coefficient, $J$ the coupling constant, $\eta_i$ a white noise such that $\langle \eta_i(t) \rangle=0$ and $\langle \eta_i(t) \eta_j(t') \rangle=\delta_{ij} \delta(t-t')$, and $T$ the temperature of the thermal bath to which the rotors are coupled (fixing $k_B=1$).
Importantly, for the classical XY model $\sigma=0$.  
The Mermin-Wagner theorem prevents in $d<2$ the emergence of long-range order (LRO) for this system. 
In dimension $d=2$, the system undergoes, below a critical $T_{KT}$, a Berezenskii-Kosterlitz-Thouless (BKT) phase transition 
associated to the unbinding of topological defects that leads, for $T<T_{KT}$, to quasi-long-range order (QLRO) \cite{Berezinskii1971,Kosterlitz1973, Kosterlitz1974}. We stress that all these results hold in (thermodynamic) equilibrium. 
%

On the other hand, the Kuramoto model (KM) is one of the pillars of synchronization theory, a central subject in dynamical systems.  
The classical KM~\cite{KuramotoOriginal} is formulated as a deterministic system of $N$ globally coupled phase oscillators or rotors, where each oscillator $i$ possesses its own natural, time invariant, frequency $\omega_i$. 
This can be expressed using Eq.~(\ref{eq:xymodel}) with $\sigma>0$,  $a_{ij}=1/N$, \,$\forall\, i,\, j$ and $T=0$. 
%
If the $N$ frequencies $\omega_i$ are drawn from a unimodal distribution, the system exhibits a phase transition, involving a collective rotation (frequency locking) and phase synchronization~\cite{KuramotoOriginal,AcebronBonilla, Sakaguchi1987}.  
This formally infinite dimension system exhibits a {second} order phase transition from order to disorder as the self-spinning strength increases~\cite{KuramotoOriginal}. 
If the oscillators lie on a regular lattice and $a_{ij}$  in Eq.~(\ref{eq:xymodel}) is restricted to the nearest (lattice) neighbors, 
as in the XY model, and $T>0$,  
the emergent, large-scale properties of the system strongly depend on the system dimensionality $d$~\cite{AcebronBonilla}. 
For $\omega_i = \omega_0$ for all $i$,  the system reduces to the XY model upon a global rotation of the reference frame, 
and thus the Mermin-Wagner theorem and BKT theory apply. 
When $\omega_i$ is drawn from a unimodal distribution as indicated above, for $d\geq5$, the system fully synchronizes. For $d=3,4$, the system is disordered in phase, but exhibits frequency locking \cite{Sakaguchi1987, hong2005collective, hong2007entrainment}.
%
In $d=2$, the system can only exhibit short-range order (SRO) \cite{RouzaireLevis,rouzairedynamics, lee2010vortices}. 
The study of defects in this context have shown that the BKT scenario no longer holds as the defects  unbind and super-diffuse when $\sigma\neq 0$  
~\cite{RouzaireLevis,rouzairedynamics, lee2010vortices}.
%
\begin{figure*}
    \centering
    \includegraphics[width=1\linewidth]{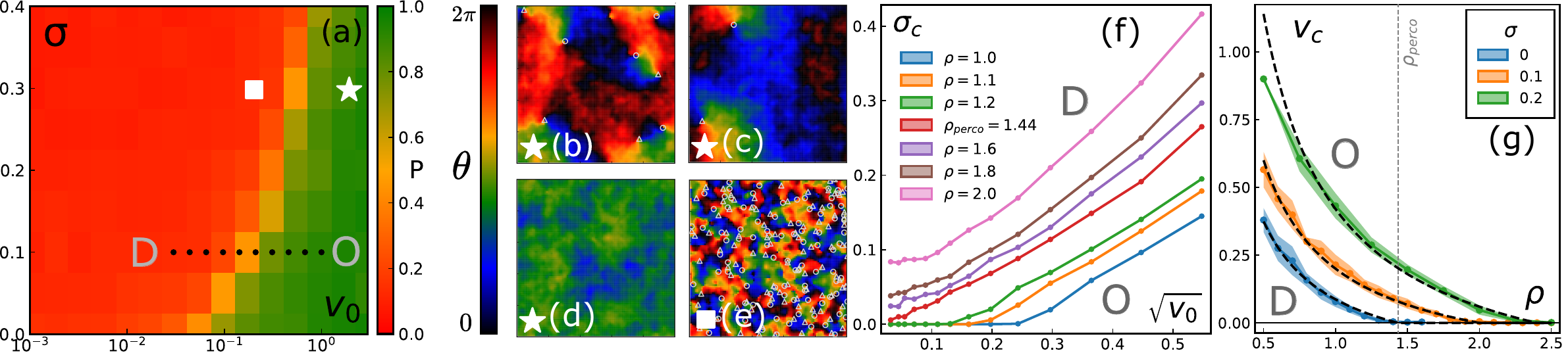}   
    \caption{\textbf{(a)} Steady polarization $P$ map in the $v_0 - \sigma$ plane for $N=10^3$, $\rho=1$.  Symbols correspond to the snapshots shown in  (b)-(e) and the horizontal dotted line to the parameters scanned in Fig.~\ref{fig:2}(a-c) across the disorder (D) to order (O) states. 
    \textbf{(b,c,d)} {Coarse-grained $\theta$ field (cf. main text and SM \cite{SM} for details)} at different times -  $50, 500$ and $1000$ - of a system with $v_0 = 2, \sigma = 0.3$ , $N=10^4, L=100$. {Circles (resp. triangles) are $+1$ (resp. $-1$) topological defects.}
      \textbf{(e)}  Steady-state {coarse-grained configuration} for $v_0 = 0.2, \sigma = 0.3$ , $N=10^4, L=100$. . 
      \textbf{(f)} Dependence of threshold $\sigma_c$ on $v_0$,  
      for $N=10^3$ and the different densities shown in the key. 
          \textbf{(g)} Critical velocity $v_c(\rho)= \alpha \big(\rho_c(\sigma)-\rho\big)/\rho$ for $N=10^4$. The black lines follow our theoretical prediction (cf. main text), with the fitting coefficients $\alpha=0.2, 0.2, 0.3$ and $\rho_c = 1.44, 2, 2.4$ for $\sigma=0,0.1,0.2$, respectively. Note that for $\rho_c(\sigma =0) = \rho_p\approx 1.44$, the percolation value, consistent with \cite{LevisPRX}. 
    }
    \label{fig:1}
\end{figure*}

 {
Although the previous paradigmatic models are restricted to a static spatial structure, synchronization patterns can also be found in a large variety of chemical and biological systems where the oscillators move in space. Some examples are fireflies moving in swarms displaying synchronized flashing~\cite{sarfati2021self}, mobile catalytic beads showing coherent oscillatory patterns of the BZ reaction~\cite{taylor2009dynamical},  amoeba exhibiting a synchronized response known as quorum sensing~\cite{gregor2010onset}, motile cells synchronizing their intracellular oscillations in somitogenesis ~\cite{oates2012patterning, uriu2014collective}, or gliding myxobacteria displaying coherent oscillations of the Frz system~\cite{guzzo2018gated}. At a theoretical level, it has been shown that in populations of identical oscillators, spatial diffusion \cite{peruani2010mobility, grossmann2016superdiffusion, LevisPRX}  promotes the emergence of order in finite systems. 
\ylann{Thus, spatial diffusivity can effectively increase  the range of interactions that, when resulting in an all-to-all coupling, leads to global order~\cite{sevilla2014synchronization, escaff2020flocking}}.  
%
However, in infinite systems spin waves cannot be suppressed for any finite diffusivity in dimension $d=1$ and thus only SRO is possible~\cite{peruani2010mobility}. While in $d=2$, the emergent order is QLRO, and the system undergoes a BKT-like transition even for diffusive oscillators~\cite{grossmann2016superdiffusion, LevisPRX}. Interestingly, if motion is super-diffusive, LRO can emerge~\cite{grossmann2016superdiffusion}. We stress that these results are restricted to populations of identical oscillators; however, fireflies, amoeba, bacteria, and organisms in general are not identical, but exhibit large inter-individual variability, even within genetically identical populations~\cite{elowitz2002stochastic, raj2008nature}. Thus, it becomes crucial to understand the synchronization of mobile and heterogeneous oscillators, where each of them possesses its own natural frequency $\omega_i$.  
Here, we address such a fundamental question using a two-dimensional model for actively moving phase oscillators, where each one has its own natural frequency. 
\ylann{The system can be considered a collection of swarmalators~\cite{okeeffe2017oscillators} where the phase dynamics is coupled with the position but the oscillator position dynamics is independent of the phase.} 
We show that, while in the absence of motion only SRO can emerge, motility allows the system to achieve QLRO and undergo a BKT transition. The relevance of these results goes beyond the realm of active matter physics.  For instance, while heterogeneous 3d static swarms of fireflies cannot exhibit synchronized flashing, our results indicate that motion can make it possible. Similarly, in somitogenesis, cellular motion might not only enhance synchronization as clearly shown in~\cite{uriu2014collective, uriu2024statistical} using identical phase oscillators, but might be essential to prevent, in heterogeneous populations, the emergence of only SRO; see~\cite{oates2012patterning, uriu2021local, fernandez2022reaction} for a realistic description of somitogenesis.


%

\textit{The Model.-- } 
 {We consider $N$ actively moving phase oscillators, at density $\rho$, such that they live in a $L\times L$ plane with periodic boundary conditions, with $L=\sqrt{N/\rho}$}. 
The phase $\theta_i$ of oscillator $i$ obeys Eq.~(\ref{eq:xymodel}) with $\omega_i$ drawn from a normal distribution, centered around $0$, of variance $1$, and where the sum over $j$ runs 
over all neighbors of $i$ defined as those oscillators at a distance less than $R_0$ and  -- i.e. 
$a_{ij}=1/{ R_0^2}$  if $||{\bf x}_j - {\bf x}_i||<R_0$, with ${\bf x}_j$ and  ${\bf x}_i$ the positions of oscillators $i$ and $j$ -- and $a_{ij}=0$ otherwise. 

The spatial dynamics of the $i$-th oscillator is given by: 
\begin{equation}
\dot{{\bf x}}_i = v_0 \, {\bf e}[\psi_i] =(v_0\,\cos\psi_i\,,\,v_0\,\sin\psi_i)
\label{eq:spatial_dynamics}
\end{equation}
where $v_0> 0$ is the  speed of the oscillator and $\psi_i$ its moving direction. 
We analyze two different scenarios: (a) ballistic motion with $\dot{\psi}_i=0$, and (b) persistent random walkers, where
$\dot{\psi}_i = \sqrt{2D_{\psi}}\,\nu_i$, with $D_{\psi}$ constant and $\nu_i$ a $\delta$-correlated noise such that $\langle \nu_i \rangle=0$ 
and $\langle \nu_i(t)  \nu_j(t')\rangle=\delta_{ij} \delta(t-t')$. 
Initially, oscillators are placed at random on the $L\times L$ plane, with $\psi_i(t=0)$  randomly selected from the interval $[0, 2\pi)$. 
%
%
In the following, we fix  $T=0.1$ and $R_0=1$, and express lengths and times in units of $R_0$ and  $\gamma/J$, {where $\gamma$ is the damping coefficient in Eq.~(\ref{eq:xymodel})}. 
Note that figures shown here correspond to the scenario (a), i.e. ballistic motion, while the same figures, but for the scenario (b), can be found in the Supplemental Material~\cite{SM}. We  obtain qualitatively the same results with both dynamics provided $v_0 D_{\psi}^{-1}$ is large enough.  
%

\begin{figure*}
    \centering
    \includegraphics[width=1\linewidth]{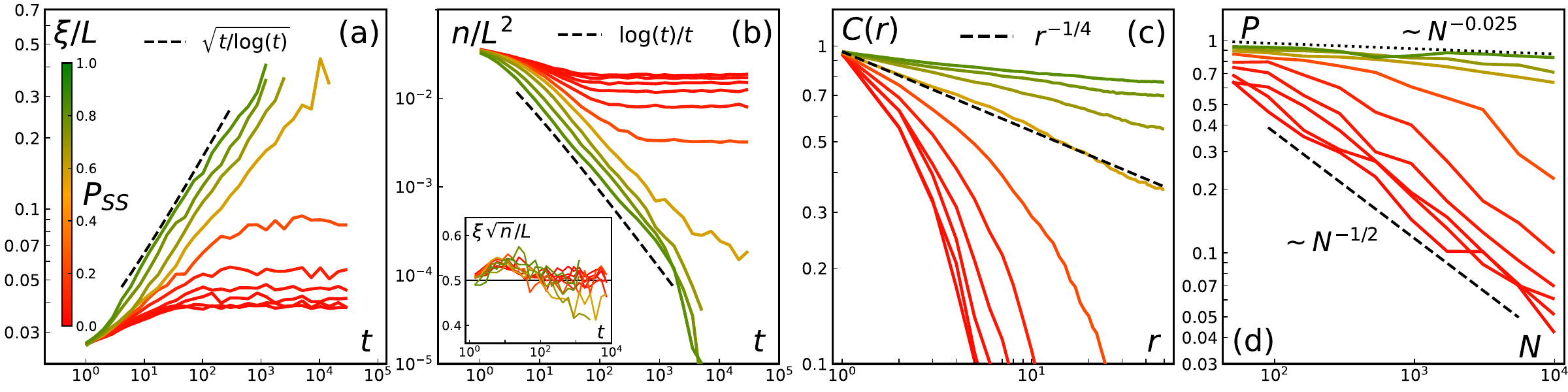}   
    \caption{
  Time evolution of:   \textbf{(a)} the spatial correlation length;  \textbf{(b)} defect density - inset: characteristic length normalized by the average distance between defects.  \textbf{(c)} Spatial decay of the correlation function in steady conditions. \textbf{(d)} Finite size scaling analysis of the polarization,   to confirming the different nature of the D and O phases. In all cases: dashed lines are XY predictions, the color of each curve corresponds to its steady  polarization value,  $N=10^4$, $\sigma = 0.1$ and $0.03 \leq v_0 \leq 1$ . }
    \label{fig:2}
\end{figure*}

\textit{Ordering in finite systems.-- }  
We characterize the emergent order using the (global) polar order parameter, defined as $P=|\sum_{j=1}^N e^{i\theta_j}|/N$. 
In parameter space $\sigma$-$v_0$, we distinguish two clear phases. 
For low  $v_0$ and high $\sigma$, the system is disordered with $P\ll1$, see Fig.~\ref{fig:1}a.  
For high $v_0$ and low-enough $\sigma$, our finite-$N$ systems reach an ordered, or synchronized, steady-state, characterized by  $P \sim 1$.  
The order-disorder {transition} in the $\sigma$-$v_0$ plane -- identified, in finite systems, by the condition $P(v_c, \sigma_c) = 1/2$, defining the thresholds $v_c$ and $\sigma_c$ -- can also be equivalently identified by monitoring the number of topological defects $n$  (see SM \cite{SM}). {To identify and track defects in our system of mobile particles, we first coarse-grain the phase field $\theta$ such that each point of an auxiliary square grid is assigned a value, as detailed in the SM \cite{SM}.
A defect is then identified as a square plaquette with a winding number $q=\pm1$, defined by the sum of the phase differences along its four bonds $\sumcirclearrowleft\Delta\theta_{i,j}=2\pi q$.}
%
Starting from a disordered initial state, polar order sets through a typical coarsening process. Topological defects initially span the entire system; they progressively annihilate two by two to eventually completely disappear: the system spontaneously breaks rotational invariance and picks one orientation, hence $P\approx 1$. 
 {We illustrate this dynamics with the coarse-grained phase fields in Fig.~\ref{fig:1}b-e. }
Occasionally, pairs of defects spontaneously generate but, in contrast to the behavior of immobile oscillators on a lattice~\cite{RouzaireLevis, rouzairedynamics, lee2010vortices}, they end up annihilating completely and the steady-state typically displays no defects, cf. Fig.~\ref{fig:1}d and \cite{SM}. This holds for the entire ordered (green) region, including the $\sigma=0$ axis, see Fig.~\ref{fig:1}a.  

Furthermore, we investigate the impact of the particle density $\rho$ on the transition line (for finite $N$) between order and disorder. For low densities, the system only orders for sufficiently high $v_0$, see Fig.~\ref{fig:1}f .  Higher densities promote synchronization because each spin interacts (in an additive fashion) with more neighbors within the fixed radius $R_0$. 
In the limit $\sigma,\,v_0\to0$, the emergence of order -- in a finite system -- requires the density to be above the percolation threshold $\rho_{p} \approx 1.44$ of (penetrable) discs of radius $R_0$, a result consistent with~\cite{LevisPRX}, where the synchronization of moving, single frequency ($\sigma=0$),  oscillators was studied. We observe that the mean coordination number corresponds to $\bar z=\pi\,\rho_{p}\,R_0^2\approx 4.5$. 
For $\rho< \rho_{p}$, there exists a threshold in velocity above which the finite system is in the  ordered region. 
To understand the scaling of the threshold speed $v_c$ with the density $\rho$ in Fig.~\ref{fig:1}g, corresponding to low densities, we estimate the number of encounters (i.e collisions, as a proxy of the number of neighbors), a particle (in the ballistic scenario) experiences in time $\tau_0$ as $z_{\text{eff}}(v_0) \propto \rho R_0 v_0 \tau_0$. Here, $\tau_0\sim J^{-1}$ is the timescale associated to the alignment torque in Eq~(\ref{eq:xymodel}). 
Requiring that $z_{\text{eff}} \simeq \bar z$, we find that $v_c \propto (\rho_c-\rho)/\rho\,$, where $\rho_c$ is a critical density increasing with $\sigma$ and satisfying $\rho_c(\sigma=0)=\rho_p$ ; see dashed lines in Fig.~\ref{fig:1}g.  

\textit{BKT transition.-- } 
To characterize the nature of the reported phases, we study the coarsening dynamics and the spatial correlations. 
Given the resemblance with a BKT scenario, we start out by providing a brief summary of key features of the classical, two-dimensional on-lattice XY model that we will use as reference.  
%
%
For $T<T_{KT}$, where $T_{KT}$ is the critical temperature, the number of topological defects $n(t)$ decreases as $n\sim \log t /t$, while the polar order parameter $P$ and the correlation length $\xi$ , defined as $C(\xi) = 1/e$, both grow over time as $\sqrt{t/\log t\,}$ \cite{YurkeHuse1993, jelic2011quench}. 
The spatial correlation function $C(r; t)$ (or spin correlation function) is defined as $C(r;t) = \langle \cos(\theta_i(t)-\theta_j(t))\rangle|_{|{\bf x}_i-{\bf x}_j|=r}$, where  $\langle\cdots \rangle$ denotes  average over thermal noise, intrinsic frequencies, initial phases, initial locations and direction of travel of individual particles.  
In the steady state, for $T < T_{KT}$, $C(r)$ decays as a power-law such that $C(r) \sim r^{-\eta(T)}$, while for $T > T_{KT}$,  $C(r)$ decays exponentially as $C(r) \sim e^{-r/\xi}$. Importantly,  at $T = T_{KT}$,  $C(r) \sim r^{-1/4}$. 
We represent these XY predictions as dashed lines in Fig.~\ref{fig:2}a-c. 

In systems of moving oscillators, we find that  the defect density $n(t)/L^2$ quickly saturates with time $t$ to a finite value 
when the system is located, on the parameter plane $(v_0, \sigma)$, in the disordered region; 
see red curves in Fig.~\ref{fig:2}. 
In this phase, the spatial correlation function decays geometrically [Fig.~\ref{fig:2}c], a signature of short-range order (SRO), 
and the correlation length $\xi$ remains always $\xi\ll L$ [Fig.~\ref{fig:2}a]. 
In contrast, in the ordered phase, topological defects progressively annihilate and  the density of defects $n(t)/L^2$ decays as 
$n \sim \log t /t$ [see green curves in Fig.~\ref{fig:2}b], while the  correlation length $\xi$ increases as $\xi \sim \sqrt{t/\log t\,}$ [see green curves in Fig.~\ref{fig:2}a]. 
The resemblance of those expressions is not a coincidence:  $\xi \sim (n(t)/L^2)^{-1/2}$ implies that the average distance between defects  controls the typical correlation length in the system. 
Moreover, a  distinctive hallmark of the XY model at low temperatures is displayed in the inset of Fig.~\ref{fig:2}b, where $\xi \sqrt{n}/L\approx 1/2$ for all $t$, 
as observed in the XY model on a lattice ~\cite{rouzairedynamics}. 
In the ordered phase, the (steady-state) correlation functions are power-laws, as displayed by the green curves in Fig.~\ref{fig:2}c. 
This is the signature of QLRO. 
Finally, at the critical point, i.e. at the boundary between the SRO and QLRO phase, the  correlation function $C(r)$ scales as $ r^{-1/4}$; see yellow curve in Fig.~\ref{fig:2}c.  
This is observed when crossing the transition either vertically or horizontally on the $(v_0, \sigma)$ plane \cite{SM}. 
In summary, comparing  the dashed with the green and yellow curves, it becomes evident that  
the analyzed out-of-equilibrium system  exhibits all the salient features of the BKT transition.
To further support the QLRO nature of the ordered phase, we performed a finite size analysis (see results Fig.~\ref{fig:2}d), finding that polar order decreases with system size as $P\sim N^{-1/2}$ in the disorder phase, while in the order phase, the scaling is $P \sim N^{-\beta}$, with $\beta<1/16$, as expected in the  XY model on a 2D lattice \cite{nelson1977momentum, frenkel1985evidence}.  



\textit{Topological defects.-- }
%
The BKT transition is characterized by the collective behavior of topological defects.
%
For $T<T_{KT}$, defects are \textit{bounded}. There exists an 
 effective $1/R$ attractive force between defects of opposite topological charges {($q$)} separated a distance $R$~\cite{YurkeHuse1993}. 
 To assess whether the same mechanism is present in the system of actively moving oscillators, we study 
 the temporal evolution of a system whose initial condition consists of {two topological defects, one with $q=+1$ and the other with $q=-1$, } separated by a distance $R(t=0)=L/2$ {(see SM \cite{SM} for more details on this defect pair initial condition). } We set parameters in the ordered phase and monitor the distance $R(t)$ between defects. 
We find that $R(t)$ decreases with $t$ (faster for larger $v_0$) as expected in the presence of an attractive force; see Fig.~\ref{fig:3}a. 
Note that the time $\tau$  for defect annihilation is itself a stochastic variable [inset, Fig.~\ref{fig:3}a]. 
It is useful to study the distance $R$ between defects with respect to the time-to-annihilation $t^*$ (instead of $t$). 
In this equivalent description, $t^*=0$ is the moment of annihilation ($R(t^*\,=\,0) = 0$) and defects move away from each other with increasing $t^*$, see  Fig.~\ref{fig:3}b. 
The interest of $t^*$ is that now the crucial part of the dynamics, when the $1/R $ force is relevant because $R$ is small, can be analyzed with averages that are meaningful and more precise \cite{YurkeHuse1993, rouzairedynamics}.  
In particular, we observe that the curves for different $v_0$ collapse by rescaling space with $v_0^{-1/2}$ and time by $v_0^{1/2}$; see inset in Fig.~\ref{fig:3}b.  
Following \cite{YurkeHuse1993}, we solve for the functional form of the interdefect separation and obtain $R(t^*)  = \exp(W(2\pi\,t^*/\mu)/2)$, where $W$ is the Lambert function (or Product Log) and $\mu$ is the 
 mobility coefficient. We use  $\mu=1/2$, as indicated in \cite{RouzaireLevis} for the  XY model, to plot the black curve in the inset of Fig.~\ref{fig:3}b. With no fitting parameter involved, the agreement between the  theoretically predicted $R(t^*)$ and simulations  is excellent. 
This shows that  topological defects in our out-of-equilibrium system effectively follow the same interaction mechanism as in the equilibrium XY scenario, which in turn explains why the QLRO nature of the ordered phase is the same in both systems. 
This annihilation mechanism has no reason to depend on the system size, which therefore strengthen our previous results on the existence of a BKT-like transition. 
 {Finally, this key result does not depend on the specific distribution of the intrinsic frequencies. 
We have compared the gaussian case to the uniform, exponential and Cauchy distributions: both the pair annihilation and the coarsening dynamics are similar in all those cases. The results even become \textit{identical} if one imposes the same variance across the different distributions, see the SM \cite{SM} for more details.}

\begin{figure}
    \centering
    \includegraphics[width=1\linewidth]{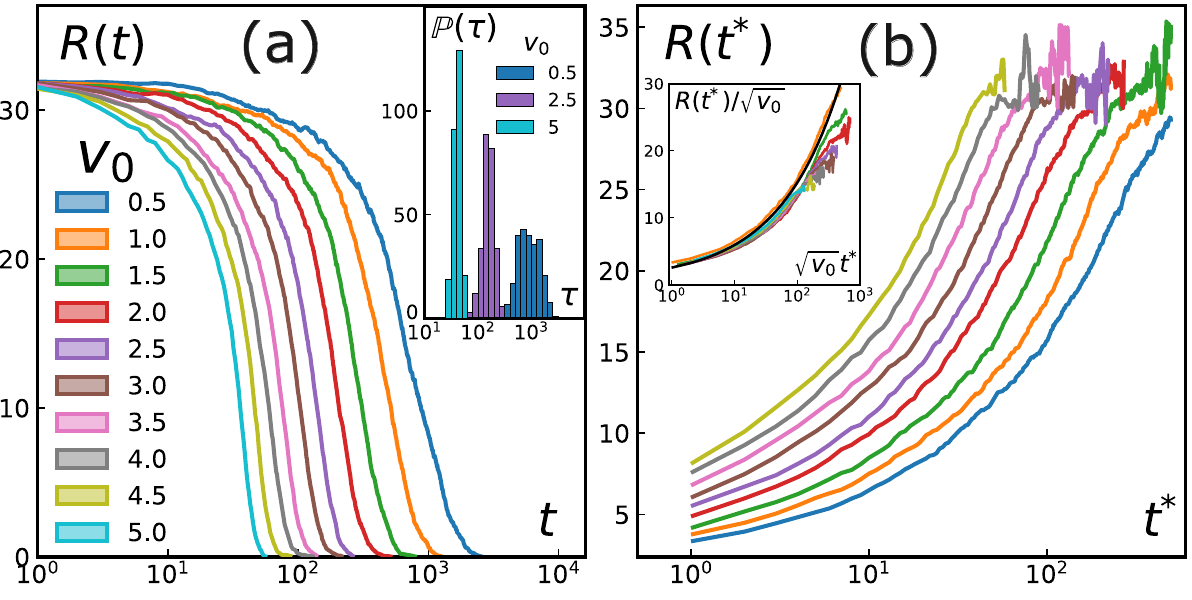}   
    \caption{\textbf{(a)} Inter-defect separation $R(t)$ for different  velocities (in the QLRO phase), $\sigma = 0.1$, averaged over 260 realizations. \underline{Inset}: Histogram of annihilation times $\tau$ for 3 selected velocities.
\textbf{(b)} Same data  plotted against the time to annihilation $t^*$. \underline{Inset}: Rescaling  time by $v_0^{1/2}$  and distance by $v_0^{-1/2}$ makes the curves collapse onto the XY predictions (in black). }
    \label{fig:3}
\end{figure}

\textit{1D spin waves in 2D.-- }
The system of moving heterogeneous oscillators exhibits some difference with respect to the classical XY model. 
In particular, we observe frequently the formation of 1D spin waves, or \textit{topologically protected state}s (TPS), characterized by a phase $\theta(x,y) = 2\pi x/L$; see Fig.~\ref{fig:4}a. These stable (or metastable) system spanning  structures form when two topological defects annihilate against the attracting force. Imagine for instance the two defects of Fig.~\ref{fig:1}c annihilating at the center of the box and not at its borders. 
We report in Fig.~\ref{fig:4}b the frequency of TPS formation, close to the order-disorder transition, where the probability to observe TPS is the highest. Such TPS patterns are (almost) not present in the classical, 2D XY model, and thus constitute a new feature of  systems of moving oscillators.

\begin{figure}
    \centering
    \includegraphics[width=1\linewidth]{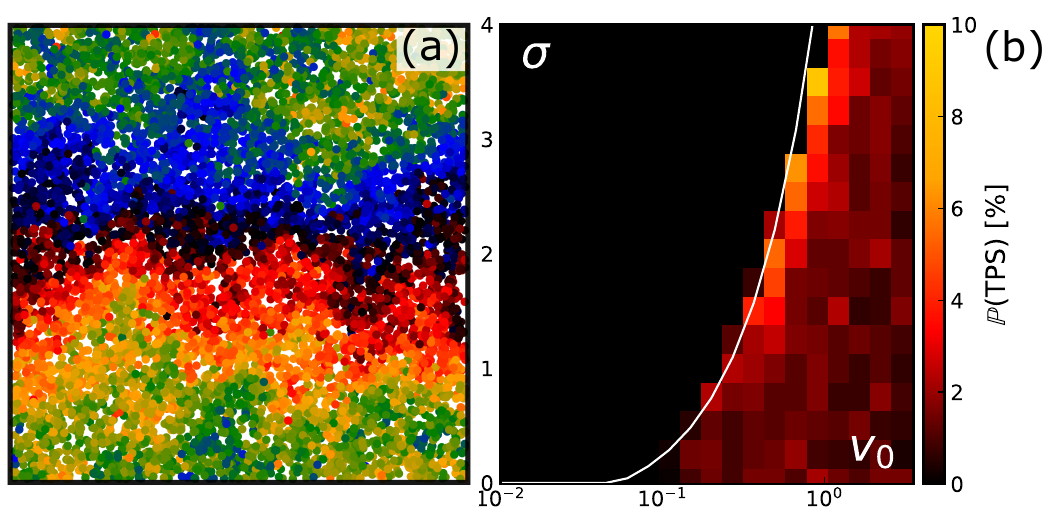}   
    \caption{\textbf{(a)} Representative topologically protected state (TPS) pattern, here for a $N=10^4, \rho=1$ system. 
\textbf{(b)}  Probability to observe a TPS against $v_0$ and $\sigma$, for $N=4000,\rho=1$, averaged over 780 realizations. Using disordered initial conditions, a TPS is identified with a defect free configuration with $P<0.5$.  
Such arbitrary threshold does not impact the results as the vast majority of the detected TPS have $P\approx 0.1$. The white line is $\sigma_c =\max(0, v_0 - 0.529) / 2 $, where the numerical factors are extracted from Fig.~\ref{fig:1}f.}
    \label{fig:4}
\end{figure}

\textit{Concluding remarks.-- } 
 {
In~\cite{peruani2010mobility, grossmann2016superdiffusion, LevisPRX} it was shown that non-equilibrium homogeneous systems of identical moving oscillators behave as equilibrium systems of (non-moving) XY spins, while
the synchronization dynamics of heterogeneous systems of moving oscillators with different natural frequencies has so far been uncharted. 
On the other hand, heterogeneous systems of static phase oscillators display only SRO in dimensions less than 5 \cite{Sakaguchi1987, hong2005collective, hong2007entrainment, RouzaireLevis,rouzairedynamics, lee2010vortices}. 
Here, we showed that in 2D, active motion makes a system of heterogeneous phase oscillators behave as an equilibrium system of static XY spins, exhibiting QLRO and a BKT transition with the same coarsening dynamics. Importantly, these results are independent of the details of the heterogeneity, i.e. whether the distribution of natural frequencies is gaussian, uniform, Laplace, or a truncated Lorentzian. 
This result connects the spatially extended Kuramoto model with the (equilibrium) XY model, and sheds light on how 
order emerges in heterogeneous active systems with quenched disorder. 
Let us recall that in active models such as the Vicsek model~\cite{vicsek1995novel}, a phase $\theta_i$ controls the  active direction of motion \cite{peruani2008mean, chepizhko2010relation, martin2018collective} of the oscillator, and the system exhibits, in 2D, LRO. 
Furthermore, such a coupling between the phase and active velocity allows heterogeneous active chiral particles to also exhibit LRO~\cite{levis2019activity}. 
Here, we have shown that active motion itself,  even in the absence of a coupling between phase and activity,  promotes the emergence of 
coordinated states that would otherwise be impossible.}

\paragraph{Acknowledgments}
D.L. and I.P. acknowledge DURSI for financial support under Project No. 2021SGR-673. 
I.P.  and Y.R. acknowledge support from Ministerio de Ciencia, Innovaci\'on y
Universidades MCIU/AEI/FEDER for financial support under
grant agreement PID2021-126570NB-100 AEI/FEDER-EU. D.L. acknowledges MCIU/AEI for financial support under
grant agreement PID2022-140407NB-C22. 
I.P. acknowledges Generalitat de Catalunya for financial support under Program Icrea Acad\`emia and project 2021SGR-673.
P.R. and F.P. acknowledge financial support from C.Y. Initiative of Excellence (grant Investissements d'Avenir ANR-16-IDEX-0008), INEX 2021 Ambition Project CollInt and Labex MME-DII, projects 2021-258 and 2021-297.

\bibliography{biblio}
\iftrue
\newpage
\appendix

\section{Supplementary Movies}


\section{Implementation details}
The code for ballistic is written in Julia language and is accessible publicly on GitHub at \href{https://github.com/yrouzaire/kuramoto_ballistic}{github.com/yrouzaire/kuramoto\_ballistic}.\\
Visit \href{https://julialang.org/}{julialang.org} to install Julia.

\subsection{Timestep for the ballistic simulations}
The timestep $dt$ used varies as a function of the simulation parameters in order to fully resolve the different phenomena at play (we indicate between square brackets which parameters it depends on): 
\begin{enumerate}
    \item The timescale for one particle to cross the interaction radii $[v_0, R_0]$ 
    \item The rotation induced by the highest intrinsic frequency (in absolute value) of the system $[N, \sigma]$.  We use the bound $\mathbb{E}\max\limits_{i} |\sigma\,\omega_i| < \sigma\sqrt{2\log N}$, inspired from the "simple answer" of Sivaraman (\href{https://math.stackexchange.com/questions/89030/expectation-of-the-maximum-of-gaussian-random-variables}{see here}) . \\
If one defines $Z = \max_{i} \omega_i$, by Jensen's inequality one obtains
\begin{align}
        \exp \{t \mathbb{E}[Z]\} &\leq \mathbb{E} \exp \{t Z\} \\
        &=\mathbb{E} \max _i \exp \left\{t \omega_i\right\} \\
        &\leq\sum_{i=1}^N \mathbb{E}\left[\exp \left\{t \omega_i\right\}\right]\\
        &=N \exp \left\{t^2 \sigma^2 / 2\right\}
        \label{eq:bound_dt_omega}
\end{align}
where the last equality follows from the definition of the Gaussian moment generating function.
Rewriting this, $\mathbb{E}[Z] \leq \frac{\log N}{t} + \frac{t \sigma^2}{2}$
Now, set $t = \frac{\sqrt{2 \log N}}{\sigma}$ to obtain $\mathbb{E}\, Z= \mathbb{E}\max\limits_{i} |\sigma\,\omega_i| < \sigma\sqrt{2\log N}$

    \item The additive alignment force $[J,\rho,R_0]$
    \item  The thermal noise $[T]$
\end{enumerate}
Concretely, 
$$dt = \min \left\{\alpha \frac{2R_0}{v_0}\ ,  \  \frac{\beta}{\sigma\sqrt{2\log N}}  \ ,  \  \frac{\beta}{J\pi\rho R_0^2}\ ,  \  \beta^2 \pi/(4T)  \right\}$$
where $\alpha = 1/10$ and $\beta = \pi/20$ are two arbitrary constants. In the range of parameters explored, the third term typically is the constraining one, leading to timesteps in the range 0.01 - 0.05. 

\subsection{Generating configurations with a pair of defects }
To study the inter-defect interaction, we manually create a pair of two oppositely charged defects. The locations of the particles $(x,y)$ are still randomly drawn from $[0,L]^2$ and their direction of motion ($\psi$) is also uniformly drawn between $0$ and $2\pi$. The field $\theta$ around a topological defect of charge $q$ and shape $\mu$ located at the origin is given by $\theta(x,y) = q\text{\,atan}(y/x) + \mu$. We set $\mu=0$ without loss of generality because the equations of motion are invariant under rotation. 
So, to create a centered pair of defects separated by distance $r_0$, the phase $\theta(x,y)$ of a spin at position $(x,y)$ is taken to be:
\begin{equation}
\begin{split}
    \theta(x,y) = & (+1)\,\text{atan}\left(\frac{y+r_0/2-L/2}{x-L/2}\right) \\
    & + (-1)\,\text{atan}\left(\frac{y-r_0/2-L/2}{x-L/2}\right)
\end{split}
\end{equation}
where $L$ is the linear size of the system.  The result is shown in Fig.~\ref{fig:manually_created_pair}.

\subsection{Smoothing procedure for visualisation and defect tracking}
The numerical identification of topological defects is complicated when particles do not live on a grid, in particular because the density is not constant neither in time nor in space, even though it remains homogeneous on average.
To overcome this issue, and also for the sake of visualisation, we coarse grain the orientation field $\theta$. 
We decompose the $L \times L$ system into a 2d grid with each cell measuring $R_0 \times R_0$. To attribute each cell an effective orientation $\tilde \theta$, we aggregate the contribution of all the particles within a circle $\mathcal{C}$ of arbitrary radius $4R_0$, centered on the considered cell. The contribution of each particle to the considered cell is weighted by an exponentially decaying function of the distance $r_k$ of the particle to the center of the cell: $\tilde \theta = \text{Arg}\left[    \sum \limits_k   \exp{(   i\theta_k - r_k/R_0)}\right]$.

\begin{figure}
    \centering
    \includegraphics[width=1\linewidth]{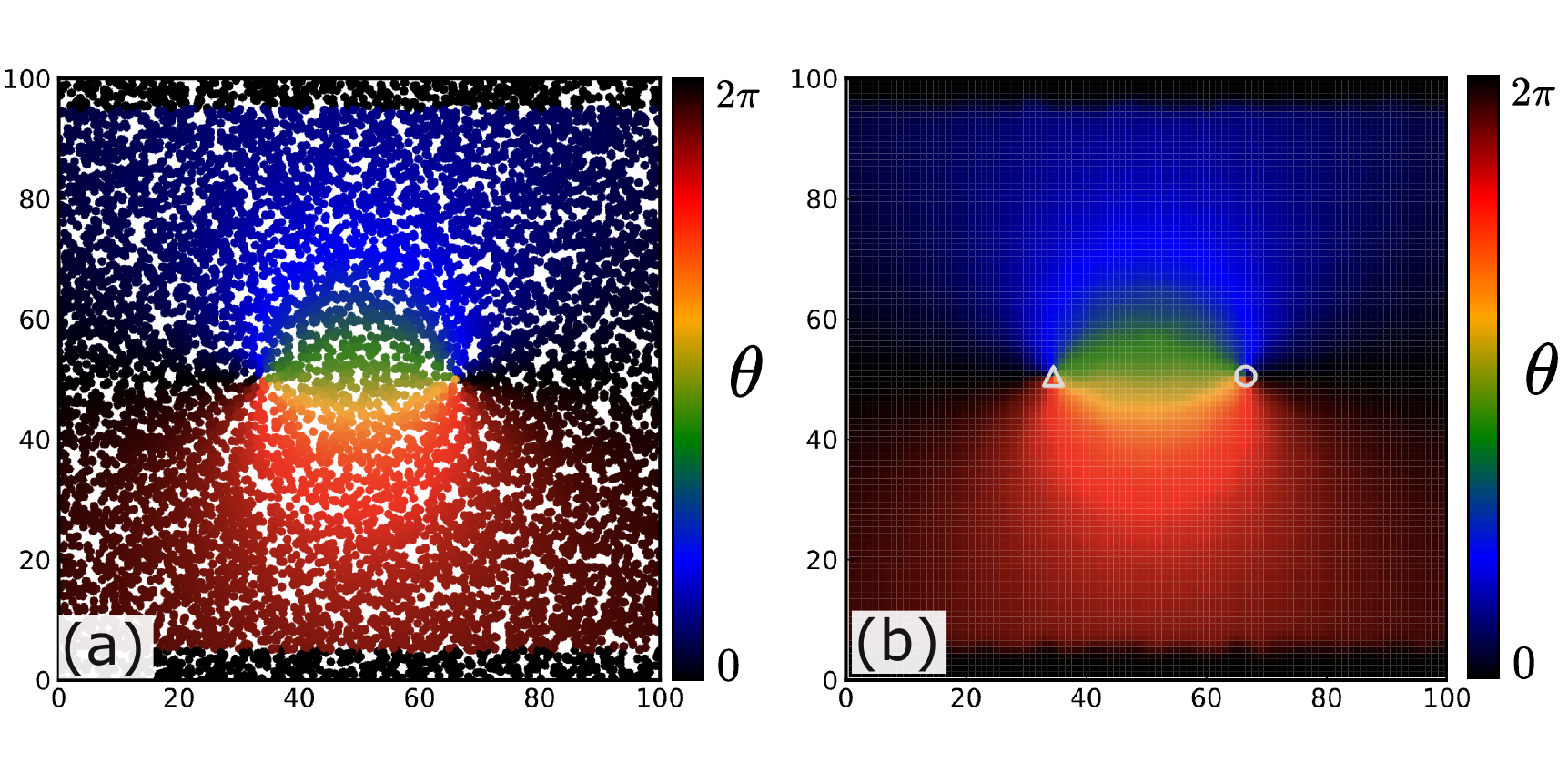}   
    \caption{A manually created pair of defects, with $q=1, N=10^4, \rho =1, L=100$, and $r_0 =32$. \textbf{(a)} Actual scatterplot of the particles.  
\textbf{(b)}  Result of the smoothing procedure. Circles (resp. triangles) are $+1$ (resp. $-1$) topological defects.}
    \label{fig:manually_created_pair}
\end{figure}

\section{Complementary results for systems of ballistic particles}

In Fig.~1(a) of the main text, we presented the phase space $v_0 - \sigma$ of the polarization $P$, for $N=10^3$ and $\rho=1$. 
In Fig. \ref{fig:phase_spaces}, we complete this picture by showing the corresponding phase space for the number of topological defects $n$ (second row) and the fluctuations in the number of defects (third row). 
We compute the ratio of the standard deviation over the mean number of defects, where both the standard deviation and the mean are calculated over 40 independent realizations of the thermal noise and initial random configurations. We compare these quantities for $\rho=1$ (below the percolation density $\rho_{perco}\approx 1.44 $, left column) and $\rho=1.9$ (above the percolation density, right column).

The number of defects $n$ and the polarization $P$ have the same trend.  Therefore, the transition can be thought either in terms of $P$, as done in the main text, or in terms of the number of defects. We highlight this duality by plotting the $P=1/2$ separation with white dashed lines, which perfectly corresponds to the transition in terms of the number of defects. 

In panels (e) and (f), we plot the fluctuations in the number of defects on a logarithmic scale. The white area is the region where the number of defects is identically zero. We see that in the high defect density region (to the left of the white dash line), the fluctuations are small. In contrast, it is at the entrance of the low defect region that the fluctuations are the largest. 

In panel (a), we also add horizontal dots and vertical triangles to highlight the values used in Fig.~\ref{fig:scan_vertical_horizontal} for the horizontal and vertical scans through the transition.  

\begin{figure}
    \centering
    \includegraphics[width=1\linewidth]{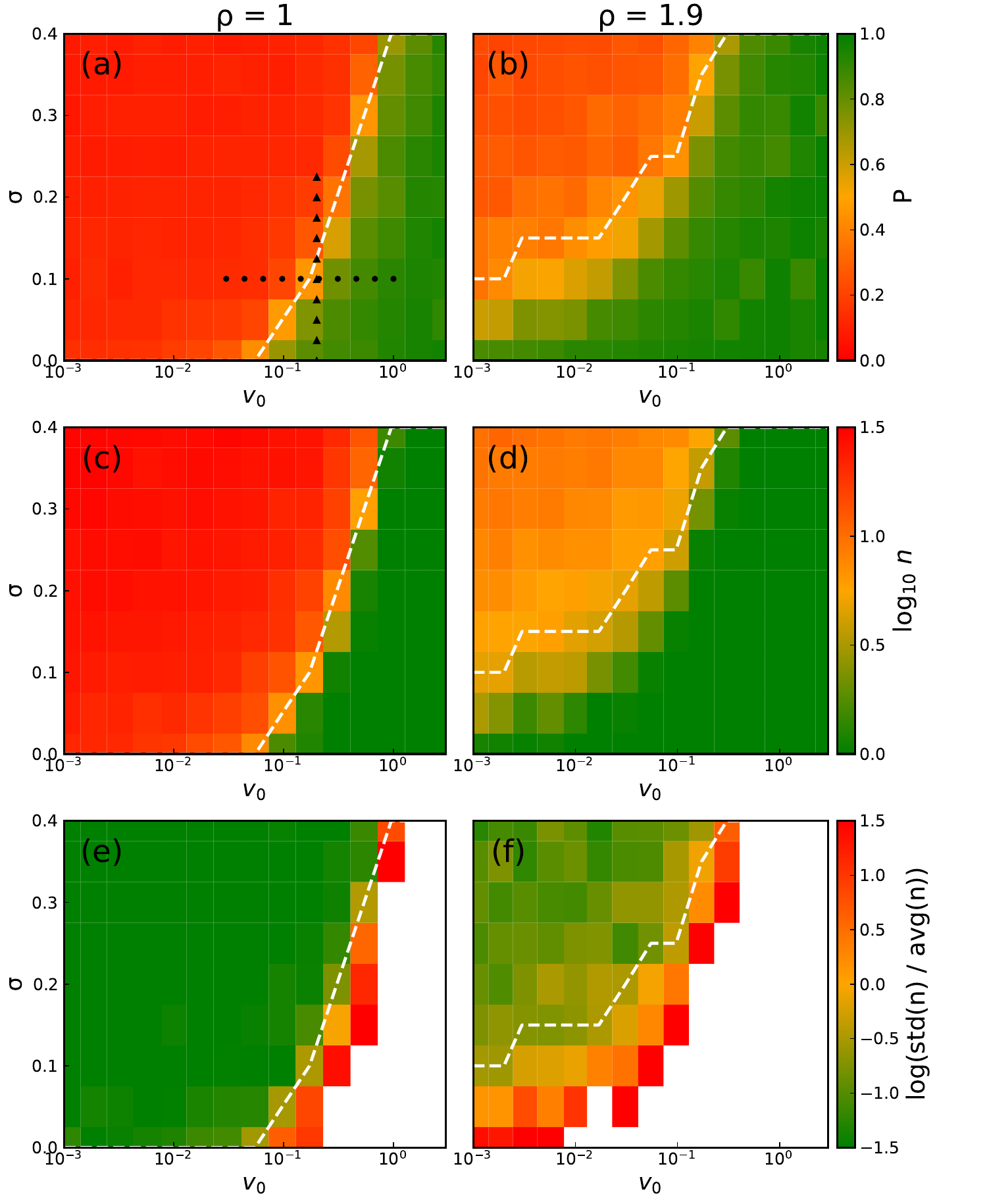}   
    \caption{Phase spaces of \textbf{(a,b)} the polar order $P$, \textbf{(c,d)} the number of defects $n$, and \textbf{(e,f)} the fluctuations of the number of defects (standard deviation divided by the mean), on a logarithmic scale.
\textbf{Left column}: At particle density $\rho=1.0$ and
\textbf{Right column}: At particle density $\rho=1.9$.
Dots mark the horizontal scan through the transition of Fig.~\ref{fig:scan_vertical_horizontal}(a-c). Triangles mark the vertical scan through the transition of Fig.~\ref{fig:scan_vertical_horizontal}(d-f)
}
\label{fig:phase_spaces}
\end{figure}

In Fig. 2 of the main text, we scanned the transition horizontally and measured the defect density, the characteristic length and the spatial correlation functions, observing how they are altered when the transition is crossed. 
In Fig.~\ref{fig:scan_vertical_horizontal}, to show that the two scans are equivalent, we compare the same quantities for both a horizontal scan (first row, panels (a,b,c), corresponding to the $\bullet$ symbols in Fig.~\ref{fig:phase_spaces}(a)), and a vertical scan (second row, panels (d,e,f), corresponding to the $\blacktriangle$ symbols in Fig.~\ref{fig:phase_spaces}(a)). 
\begin{figure}
    \centering
    \includegraphics[width=1\linewidth]{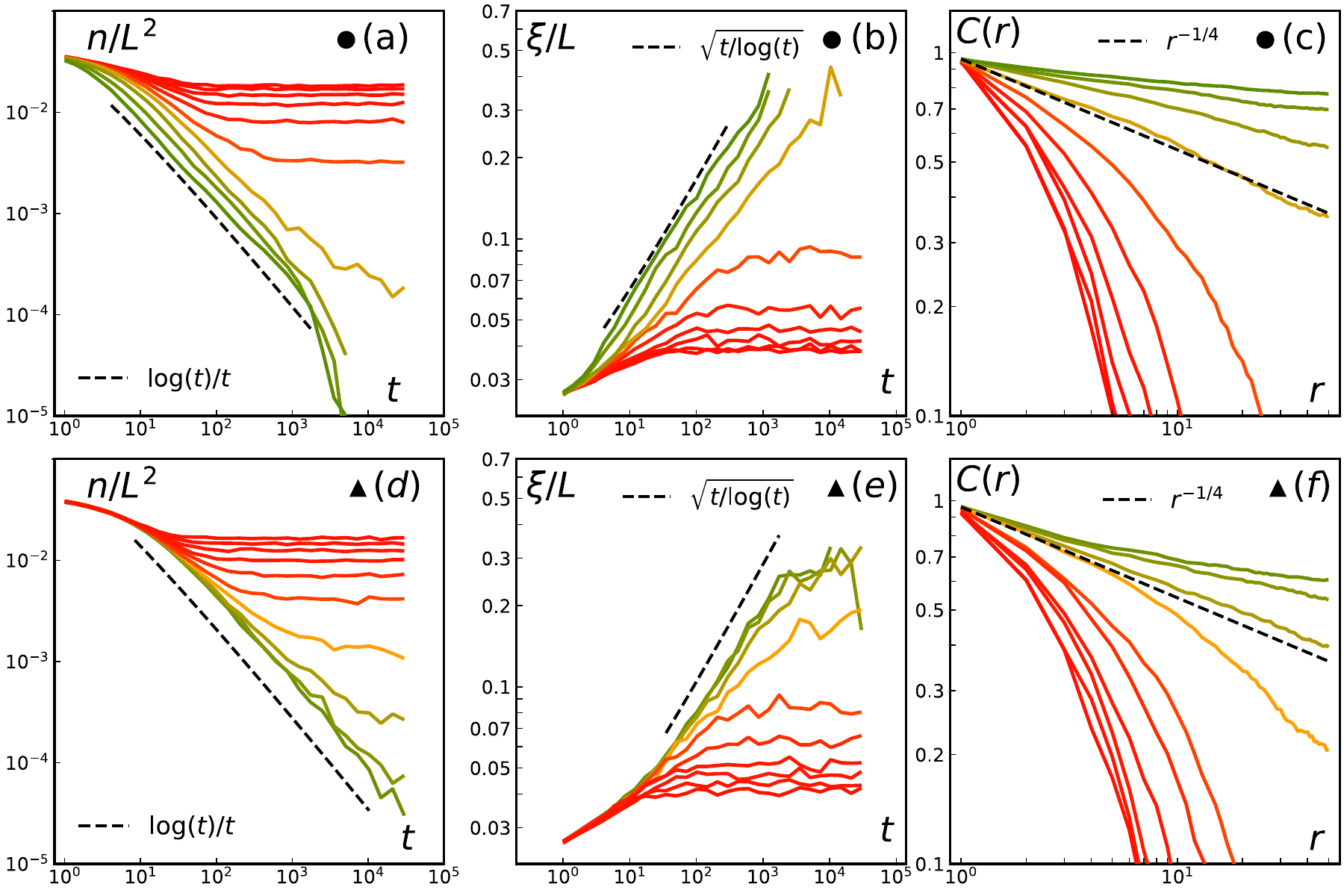}   
    \caption{
    Measurements on a system with $N=10^4$ particles.
Each curve is colored by its corresponding steady-state polarization.
For the horizontal scan \textbf{(top row)}, $\sigma = 0.1$ and $0.03 \leq v_0 \leq 1$.  For the vertical scan \textbf{(bottom row)}, $v_0 = 0.2$ and $0.025 \leq \sigma \leq 0.225$.   \textbf{Left column}: defect density over time.\textbf{ Central column}: normalised characteristic length over time.\textbf{ Right column}: Spatial correlation functions in the steady state. 
\\
}
\label{fig:scan_vertical_horizontal}
\end{figure}

In the left panel of Fig.~\ref{fig:relaxation_time}, we plot the polarization $P$ as a function of the number of particles $N = 50, ..., 40000$, averaged over 40 independent realizations, for $v_0 = 1$ and $\sigma = 0.1$ (in the quasi-ordered phase). The different colors represent different times, exponentially spaced, from bottom/blue to top/brown. At $t=4\times 10^4$ (in brown, below the dotted line), the systems up to $N=40000$ have relaxed. We see that smaller systems relax faster than bigger ones. In the right panel of the same Fig.~\ref{fig:relaxation_time}, we report the relaxation time the system needs to relax to its steady state in terms of the polarization. We define this time $t_{relax}$ as the time needed for the system to reach $95\%$ of its steady state polarization:  $P(t_{relax})=0.95\,P(t \to \infty)$. It appears that $t_{relax}$ follows the relationship $t_{relax} \sim N\,\log N$. 

\begin{figure}
    \centering
    \includegraphics[width=1\linewidth]{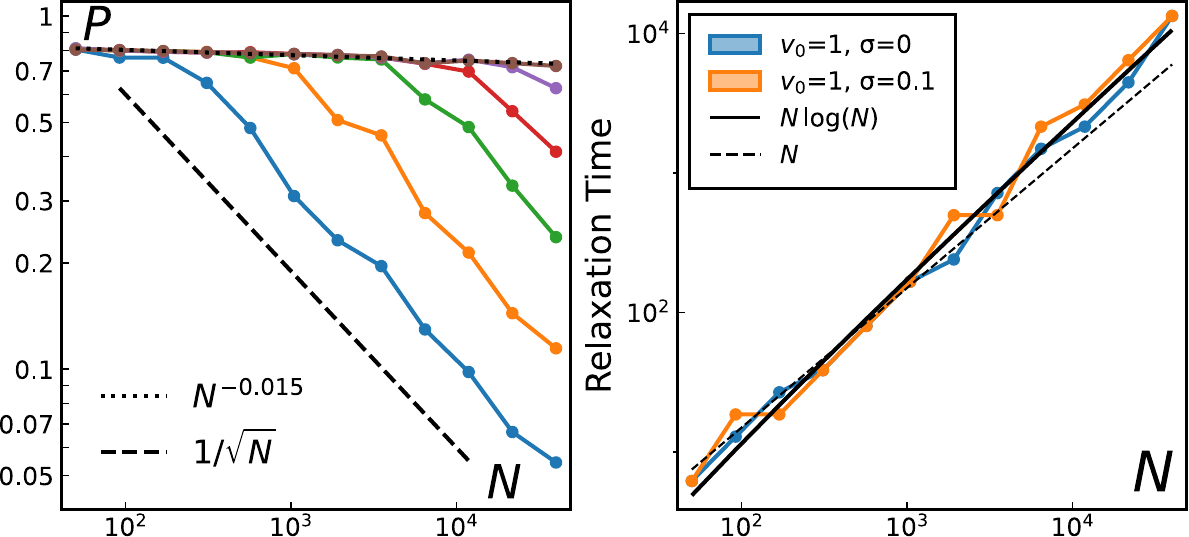}   
    \caption{\textbf{(left)} Finite size scaling of the polarization $P$ against $N$, for $v_0 = 1, \sigma = 0.1$, for different times $t=27, 166, 1035, ..., 40\,000$. (exponentially spaced, from bottom/blue to top/brown).
    \textbf{(right)} Relaxation time, defined as the time necessary to reach 95\% of the final polarization value, for different number of particles $N$. The solid black line is $0.025\,N\log N$. The dashed black line is $0.15\,N$.     }
    \label{fig:relaxation_time}
\end{figure}

 {

\subsection{The case $T=0$}
In this section, we address the specific case of the absence of thermal noise: $T=0$, as it is closer to the original Kuramoto model \cite{KuramotoOriginal} and the role of noise in the BKT topological phase transition is crucial.  
In the absence of mobility, when the spins live on a regular lattice, the $T=0$ case is somehow different from the small (but finite) temperature case. Indeed, the topological defects are prone to getting trapped in specific regions of space, where the configuration of angles $\theta$ and intrinsic frequencies $\omega$ make them cycle in space roughly periodically, destroying ergodicity. 
In the present work, the mobility of individual agents implies a constant change in the neighborhood of each rotor, providing a non-thermal source of fluctuations. This is why we observe qualitatively similar behaviors for $T=0$ and $T>0$. This is true for the defect pair annihilation process and for the coarsening dynamics.
Figure~\ref{fig:comparison_T0}(a) shows that the distance between two manually created defects decreases in the same fashion for $T>0$ (dashed lines) and $T=0$ (solid lines). 
It also shows that the characteristic lengthscale $\xi$ (panel b) and the defect density $n/L^2$ (panel c) both exhibit the same behavior across the order-disorder transition. In the disordered region of phase space, $\xi$ and $n/L^2$ saturate at a final value. In the ordered region, $n/L^2$ decays to 0 following the same scaling as the one of the XY model (dashed lines), allowing for the typical lengthscale to grow up to the system size. 

The subtlety of the zero temperature case resides in the nature of the order. While for finite temperatures, the spatial correlation function $C(r)$ decays with a powerlow, indicating QLRO (cf. Fig.~2c of the main text), this is no longer the case for the $T=0$ case. When the system is no longer coupled to a thermal bath, the $C(r)$ decays faster than algebraically, indicating true long range order (LRO), see. Fig.~\ref{fig:comparison_T0}(d).

\begin{figure}
    \centering
    \includegraphics[width=1\linewidth]{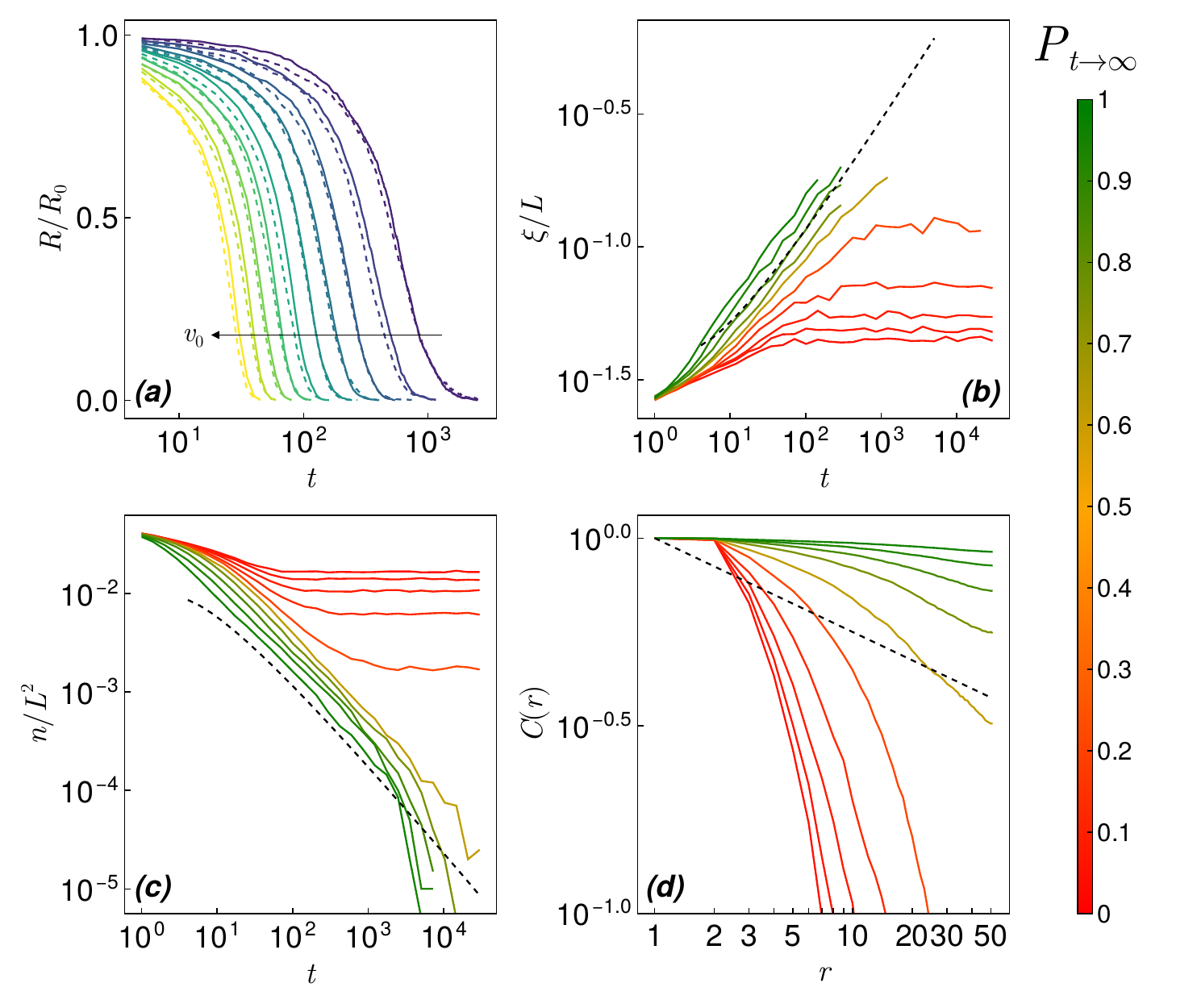}   
    \caption{ {\textbf{(a)} Distance between 2 defects manually created $R(t)$, for $\sigma=0.1$, different $v_0 = 0.5, 1, 1.5, ..., 5$. In dash, $T=0$, in solid line $T=0.1$. Equivalent to Figure 3(a) in the main text.
    \textbf{(b-d)} Coarsening dynamics for the $T=0$ case, for various parameters $v_0, \sigma$ crossing the transition horizontally (circles in Fig.~\ref{fig:phase_spaces}a) . The characteristic lengthscale $\xi(t)/L$ in panel (b), the defect density $n(t)/L^2$ in panel (c) and the spatial correlation functions $C(r)$ in panel (d). Equivalent to Figure 2(a-c) in the main text.}}
    \label{fig:comparison_T0}
\end{figure}

\subsection{Different distributions of the intrinsic frequencies $\omega$}
In this section, we show the robustness of all the previous results to substantial changes in the distribution of the intrinsic frequencies $\omega$. In the main text, all the results are for a Gaussian distribution : $\omega_i \sim \mathcal{N}(0, \sigma^2)$. Here, we compare to broader distributions: the Laplace distribution (a symmetrized version of the exponential distribution) and the uniform distribution. To make ties with the original Kuramoto model, we also include the fat-tail Cauchy(= Lorentzian) distribution. Since the Cauchy distribution decays like $x^{-2}$ for large $x$, none of the moments is defined. We thus truncate it to an arbitrary threshold $\pm 5\,\sigma$, where here $\sigma$ is the width at mid-height (called the scale parameter). 
For all those very different distributions, the system exhibits \textit{qualitatively} similar dynamics. Even better, if one matches the variance of all these distributions, the results are \textit{quantitatively} the same, see Fig.~\ref{fig:distribution_omegas}(a,b) for the pair annihilation process. 

By "matching the variances", we mean that if $\sigma^2$ is the variance of the Gaussian distribution $\mathcal{N}(0, \sigma^2)$, then we compare with a centered Laplace distribution with probability density $\exp \left( -\frac{|x|}{\sigma/\sqrt{2}} \right)/(\sqrt{2}\sigma)$ and with a uniform distribution between $-\sigma \sqrt{3}$ and $+\sigma \sqrt{3}$. For the truncated Cauchy, we take the scale parameter equal to $\sigma$, such that the probability density is $\left(\pi \sigma\left[1 + \left(\frac{x-x_0}{\sigma}\right)^2\right]\right)^{-1}$  . 
Note that here we do not use the criteria  $\mathbb{E}\max\limits_{i} |\sigma\,\omega_i| < \sigma\sqrt{2\log N}$ (cf. the first paragraph of the SM) to determine the timestep $dt$ anymore, as the inequality is valid for a Gaussian distribution only. We simply take $dt = \beta / \min\limits_i \{\omega_i\}$, where $\beta = \pi/20$ is an arbitrary constant. This ensures that the oscillator of the highest frequency is well resolved. This also motivates the truncation of the Cauchy distribution; indeed, a genuine Cauchy distribution, because even its first moment is not defined, will lead to $\max\limits_{i} |\omega_i| \to \infty $, and thus $dt \to 0$ in the thermodynamic limit $N\to\infty$. 

Panels (a) and (b) indicate that the annihilation dynamics of a pair of defect is independent of the specific distribution one sample the intrinsic frequencies from. We have also checked for the uniform distribution (with $\sigma =0.1$) that the large scale dynamics follows the one described in the main text. In Fig.~\ref{fig:distribution_omegas}(c), we indeed report the the spatial correlation functions $C(r)$ in the steady-state exhibit the same qualitative behavior than the base Gaussian case, indicating that the system still develops quasi-long range order. Our results are thus very robust and general, underlying the fundamental role of mobility in the topological phase transition in a heterogeneous population of oscillators. 

} 

\begin{figure*}
    \centering
    \includegraphics[width=1\linewidth]{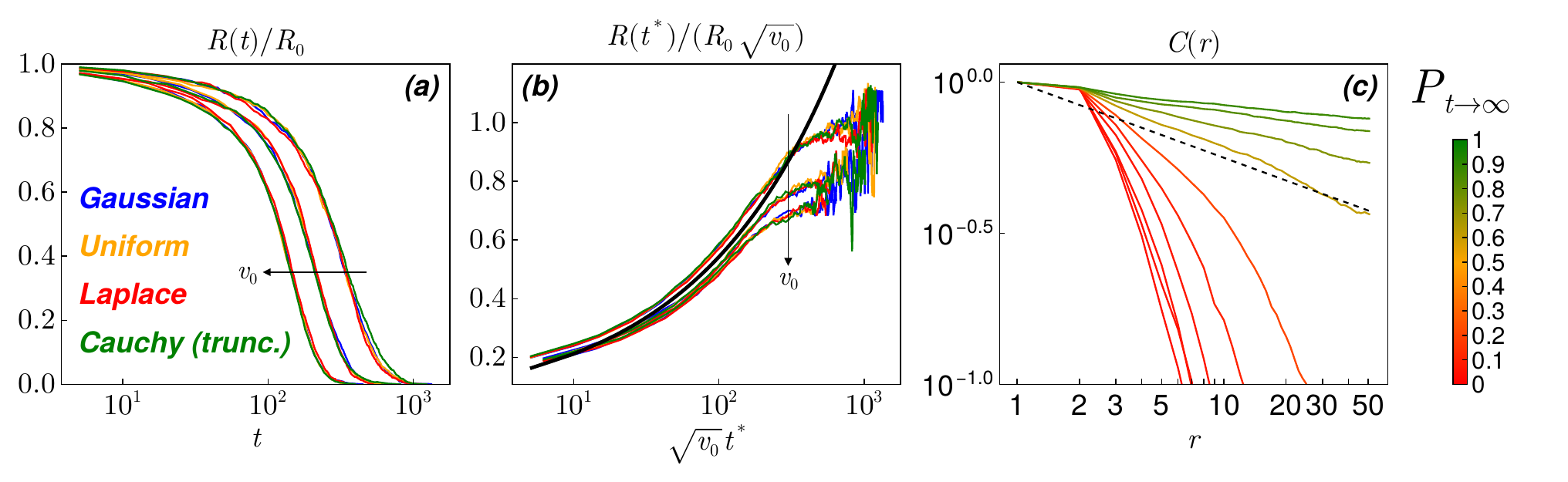}   
    \caption{ {\textbf{(a)} Decay of the distance between two defects manually created at an initial distance $R_0$, for $T=0.1$. Results are averaged over 400 independent realizations. Each group of lines is a different $v_0 = 1, 1.5, 2$ from right to left. Each color is from a different distribution of the $\omega$. For each distribution, we keep the same standard deviation $\sigma=0.1$ (see text for details). 
    \textbf{(b)}     Same data but on rescaled axes. $t^*$ is the time to annihilation and the black line is the XY prediction, computed from the results of \cite{YurkeHuse1993} (both explained in more detail in the main text).  \textbf{(c)} Spatial correlation functions $C(r, t\to \infty)$ for $T=\sigma=0.1$ for the \textit{uniform distribution}. Each curve corresponds to a given velocity $v_0$ (crossing the transition, see circles in Figure.~\ref{fig:phase_spaces}a) and is colored according to the polarization $P$ in the steady-state (see colorbar).}
    }
    \label{fig:distribution_omegas}
\end{figure*}

\section{Results for systems of persistent random walkers}
In order to obtain a phase space plot for the behavior of diffusive particles, equivalent to the one for ballistic particles, i.e. in terms of $\sigma - v_0$, we need to consider the spatial dynamics of diffusive particles:

\begin{equation}
 \dot{x_i} = v_0 \begin{pmatrix}  \cos \psi_i \\ \sin \psi_i  \end{pmatrix} \\ , \dot{\psi_i} = \sqrt{2D_{\psi}}\,\nu_i 
\end{equation}
where $\dot{x_i}$ represents the translational motion of particle i with a constant speed $v_0$, and $\dot{\psi_i}$ represents its rotational motion with a rotational diffusion coefficient $D_{\psi}$. Persistence length defined as $l_p=v_0 /D_{\psi}$, is a characteristic length scale over which angular correlations decay. Scanning the behavior of diffusive systems for different $v_0$ values affects the value of $l_p$. Therefore, we are left with two possible procedures:
Keeping $l_p$ constant, therefore, in addition to $v_0$ we also need to change $D_{\psi}$. In this case, spatial diffusion also changes: $D_{spatial}=v_0^2/(2D_{\psi})=l_pv_0/2$. The other option is to keep $D_{spatial}$ constant, therefore changing $l_p$. In all parts of our study, we keep persistent length constant at $l_p=1$. With this choice, we present the most essential results, similar to those obtained for ballistic particles.

Fig.~\ref{fig:HeatMap_pers} (a) shows the phase transition of a system of persistent random walkers for particle density $\rho=1$. For small $v_0$ and large $\sigma$ the system is disordered, while for larger $v_0$ values, and smaller $\sigma$ values, order emerges in our finite-$N$ systems. The effect of particle density on the transition from disorder to the ordered phase is shown in panels (b) and (c). Panel (d) shows the percentage of occurrence of 1D spin waves (TPS) in the system of persistent random walkers. It can be observed that the general results are consistent with those obtained in the main text for ballistic particles.

Finite size scaling analysis of the diffusive system supports quasi long-ranged order. Polarization $P$ scales as $P \sim N^{-\beta}$, and $\beta<1/16$, a behavior observed for the XY system; Fig.~\ref{fig:hov} (a, b) \cite{nelson1977momentum, frenkel1985evidence}.
Fig.~\ref{fig:hov} (c, d) shows the behavior of the correlation function along the specified regions of phase space shown in Fig.~\ref{fig:HeatMap_pers} (a). 
The dynamics of defect separation are illustrated in Fig.~\ref{fig:defects}, where you can observe that the distance between two manually created defects decreases with time, indicating an effective attraction force between the defects.\\ 
\ \\
NB: Figures \ref{fig:HeatMap_pers}, \ref{fig:hov} and \ref{fig:defects} are on the pages below. 
\onecolumngrid

\begin{figure}[]
    \centering
    \includegraphics[width=0.5\linewidth]{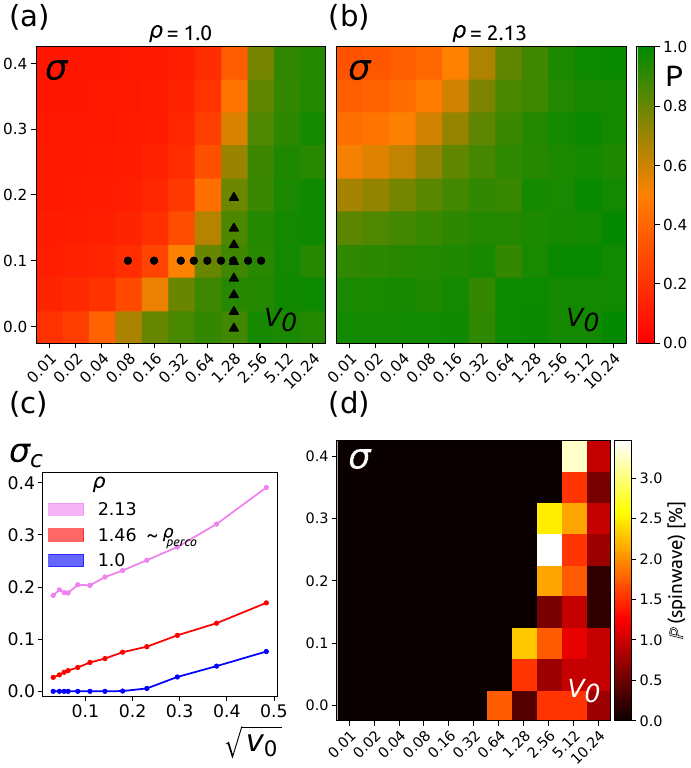}   
    \caption{\textbf{(a), (b)} Phase space of a system of persistent random walkers for polarization $P$ across the $\sigma-v_0$ parameter space for two values of density. The horizontally and vertically placed markers in \textbf{(a)} indicate the parameter sets used in Fig.\ref{fig:hov} for exploring system properties through the transition region. System parameters include $N=1225$ and $L= 35, 24$, corresponding to densities $\rho=1.0, 2.13$ respectively. The polarization $P$ is computed by averaging over 40 realizations. \textbf{(c)} Variation of critical intrinsic frequency ($\sigma_c$) versus velocity ($v_0$) obtained at three different particle densities, including a density near the percolation threshold, $\rho=1.46$; system size here is $N=1225$, and the number of realizations 40. \textbf{(d)} Percentage of observing a TPS for different $\sigma$ and $v_0$ values, with $N=3969$, and $\rho=1$, obtained from $520$ realizations. 1D spin waves are detected in steady state when the absolute difference between the mean polarization $P$, averaged over realizations, and the polarization $P_r$ from a single realization, exceeds a certain threshold ($|P-P_r|>0.4$). Black cells on the left side of phase space contain no statistics.}
    \label{fig:HeatMap_pers}
\end{figure}
\begin{figure}
    \centering
    \includegraphics[width=0.6\linewidth]{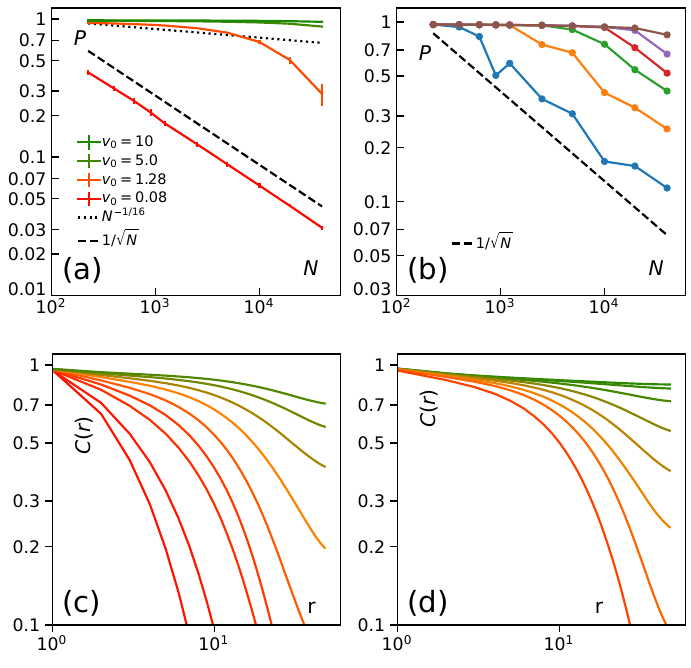} 
    \caption{Order-disorder transition properties for a system of persistent random walkers. \textbf{(a)} Finite size scaling analysis of polarization $P(N)$ for fixed $\sigma =0.1$ and varying $v_0$, indicated by the horizontally placed markers in Fig. \ref{fig:HeatMap_pers}. Colors represent the steady-state polarization of the largest system, with data averaged over 20 realizations. In the disordered phase, the order scales as $P\sim N^{-1/2}$, while in order phase, it decays with an exponent smaller than $1/16$, a signature of QLR order. \textbf{(b)} Finite size scaling analysis of polarization for $\sigma=0.1$ and $v_0=5.0$ (the middle curve in \textbf{(a)}) at different times $t=20, 120, ..., 4\,000$, starting from an initial disordered state (blue line) and transitioning to a stationary state (brown line). Dashed line shows the scaling of $P$ when the system is disordered. \textbf{(c)}, \textbf{(d)} Correlation functions obtained for the parameters along the horizontally and vertically placed markers in Fig. \ref{fig:HeatMap_pers}, respectively. The color coding corresponds to the steady-state polarization. Here, $N=10000$, $\rho=1$ with 40 realizations per parameter set. }  
    \label{fig:hov}
\end{figure}

\begin{figure}
    \centering
    \includegraphics[width=0.6\linewidth]{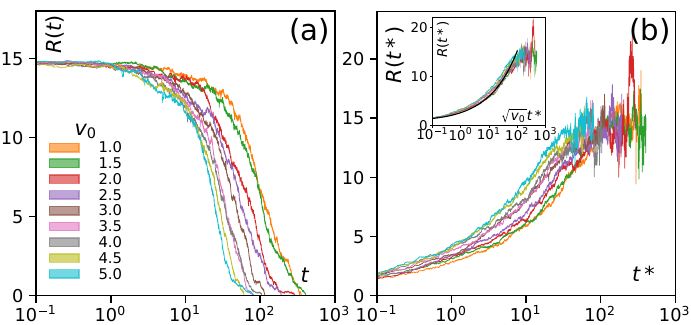} 
    \caption{Defect dynamics in a system of persistent random walkers. \textbf{(a)} Interdefect separation $R(t)$ vs time $t$ for different particle velocities (in the QLRO phase), $\sigma = 0.1$, averaged over 40 realizations. 
\textbf{(b)} The same data (and same colors) plotted against the reversed time $t^*$ (time to annihilation). \underline{Inset}: Rescaling the x-axis by $v_0^{1/2}$ make the curves collapse onto the XY predictions (in black, see the main text for equation).} 
    \label{fig:defects}
\end{figure}

\fi 
\end{document}